\documentclass{elsart}
\usepackage{natbib}
\input epsf

\newcommand{\dofig}[4]
{\begin{figure}[tbp]
\begin{center}
\leavevmode
\hbox{%
\epsfxsize=#2
\epsfbox{#1}}
\caption{#3}
\label{#4}
\end{center}
\end{figure}}
\def \xvec {\vec{x}}

\def \mhat {\hat{m}}
\def \bigT {T}
\def \vvec {\vec{v}}
\def \bigSigma {\Sigma}

\def \bigK {K}
\def \info {{\mathcal I}}
\def \avec {\vec{\alpha}}
\def \alphaPDE {$\vec{\alpha}$PDE}
\begin{document}
\begin{frontmatter}
\title{\alphaPDE:  A New Multivariate Technique for Parameter Estimation}
\author[Chicago]{B. Knuteson}
\author[Rice]{H. E. Miettinen}
\author[Finland]{L. Holmstr\"{o}m}
\address[Chicago]{Enrico Fermi Institute, University of Chicago}
\address[Rice]{Department of Physics and Astronomy, Rice University}
\address[Finland]{Rolf Nevanlinna Institute, University of Helsinki, Finland}
\begin{abstract}
We present \alphaPDE, a new multivariate analysis technique for parameter estimation. The method is based on a direct construction of joint probability densities of known variables and the parameters to be estimated. We show how posterior densities and best-value estimates are then obtained for the parameters of interest by a straightforward manipulation of these densities. The method is essentially non-parametric and allows for an intuitive graphical interpretation. We illustrate the method by outlining how it can be used to estimate the mass of the top quark, and we explain how the method is applied to an ensemble of events containing background.
\end{abstract}
\end{frontmatter}

\section{Introduction}

In an earlier paper~\cite{Holmstrom} we introduced the PDE (Probability Density Estimation) method, an essentially non-parametric and multivariate method designed for identifying small signals among large backgrounds. The method makes use of kernel density estimates for signal and background probability densities, and a simple discriminant function is then used to classify candidate events. The PDE method was applied successfully to the search for the top quark at the Fermilab Tevatron, and it is an integral part of a general search strategy~\cite{QuaeroPRL} for analyzing data from high-energy physics experiments.

In this paper we present \alphaPDE, an extension of the PDE method designed for parameter estimation, where $\avec$ represents a vector of parameters to be estimated. In many applications $\avec$ is a single parameter, such as the mass of an unstable particle detected through its decay products. This non-parametric and multivariate method may be particularly applicable to problems such as determining the mass of the top quark in the upcoming collider run (Run II) of the Fermilab Tevatron.

Multivariate methods are now widely recognized as being more powerful than 
univariate methods, and a non-parametric method has the advantage that one need not make assumptions about the forms of probability distributions.  Those who feel uneasy about the ``black-box'' quality of neural networks should welcome the straightforward manipulation of probability densities used in this method, and the intuitive graphical interpretation that results.  Because probability densities are constructed and manipulated directly, obtaining any additional statistical information --- Bayesian credible intervals, for example --- is a straightforward exercise.

A typical parameter estimation problem is described in Sec.~\ref{sec:Problem}; our recipe for solving it is provided in Sec.~\ref{sec:Recipe}.  The salient features of the method and its potential advantages are summarized in Sec.~\ref{sec:Conclusions}.

\section{The problem}
\label{sec:Problem}

The next decade of high energy collider physics will emphasize measurements and searches for new phenomena at the scale of several hundred GeV.  The existence of a new particle at this scale can be convincingly demonstrated by observing a peak in an invariant mass distribution, but the signature may be such that more indirect methods of establishing the particle's existence, and subsequently measuring parameters such as its mass and couplings, are required.  We introduce \alphaPDE\ with an example of this nature: the determination of the top quark mass.  Top quarks are pair-produced at the Fermilab Tevatron, each decaying promptly to a $W$ boson and a $b$ quark.  Each $W$ boson in turn decays either to a charged lepton and a neutrino, or to two quarks.  Quarks hadronize, appearing in the detector as collimated flows of energy (jets).  The characteristic experimental signature for a top quark event is therefore a final state containing either an energetic lepton, missing transverse energy, and several energetic jets, or a final state containing two energetic leptons, missing transverse energy, and a pair of jets; decays to six jets are difficult to distinguish from events in which no top quark was produced.  The application of selection criteria favoring events with jets originating from $b$ quarks enhances the fraction of top quark events in the sample.  

For the sake of simplicity we assume that two variables $\xvec = (x,y)$ have been identified for this analysis.  This pair might be the transverse energies of the lepton and the leading jet; it might be the invariant mass of the sub-leading jets and the transverse momentum of the $W$ boson; it might be the scalar sum of all jet transverse energies and the output of a neural network built with event-shape variables.  No special assumptions about the nature of these variables need be made.

\section{The Recipe}
\label{sec:Recipe}

The goal is to construct a method that performs as well as (or better than) such popular algorithms as neural networks, but to keep the method sufficiently simple that it reads like a recipe.  The recipe follows.

\subsection{Specify $p(m)$}
\label{sec:prior}

This method has its roots in Bayesian statistics, and as a result it has the advantage (disadvantage) of enabling (requiring) the specification of a function $p(m|\info)$ that encodes prior beliefs about the value of the top quark mass $m$.  $\info$ here is used in standard Bayesian notation to represent all assumptions implicit in our specification of this prior probability.  The basic assumptions contained in $\info$ will not change, so we drop it from here on, writing simply $p(m)$.  A natural choice for $p(m)$, used when there is strong belief that the true mass must lie somewhere between $a$ and $b$ but no reason to prefer any value within that range over any other, is the flat prior:  $p(m)=\frac{1}{b-a}$ for $a<m<b$, and $0$ elsewhere.

\subsection{Generate Monte Carlo events}
\label{sec:generateMonteCarlo}

Monte Carlo events are generated with top quark masses $m$ pulled from the distribution $p(m)$ specified above.  That is, the probability that an event with a top quark mass between $m$ and $m+\delta m$ is generated is $p(m) \, \delta m$. For each Monte Carlo event we calculate the two variables $\xvec=(x,y)$.  

A histogram in $(x, y, m)$ filled with the generated events approximates the joint density $p(x, y, m)$.  This function has the property that, given an event in which a top quark is produced and decays to the observed final state, the probability that the top quark mass was between $m$ and $m+\delta m$, the first variable between $x$ and $x+\delta x$, and the second variable between $y$ and $y+\delta y$, is simply $p(x,y,m) \, \delta x \, \delta y \, \delta m$.

\subsection{Construct a training array $\bigT$}

Each of the $N$ Monte Carlo events just generated is characterized by three numbers:  the value of $x$, the value of $y$, and the top quark mass $m$.   The Monte Carlo events are labeled with the index $i$ ($i=1,..,N$);  the three numbers corresponding to the $i^{\rm th}$ event are then $x_i$, $y_i$, and $m_i$.  Define the {\em event vector} $\vvec_i$ for the $i^{\rm th}$ Monte Carlo event by
\begin{equation}
\vvec_i = (\xvec_i,m_i),
\end{equation}
and define the {\em training array} $\bigT$ for the entire set of Monte Carlo events by
\begin{equation}
\bigT_{ij} = (\vvec_i)_j.
\end{equation}
Here and below $i$ ranges from 1 to $N$ and indexes the Monte Carlo events; $j$ ranges from 1 to 3 and indexes the components of the event vector $\vvec$.

\subsection{Calculate the covariance matrix}

Having defined the event vector $\vvec$, calculate the {\em mean event 
vector}
\begin{equation}
\langle \vvec \rangle= \frac{1}{N} \sum_{i=1}^{N}{ \vvec_i }
\end{equation}
and construct the {\em training covariance matrix} 
\begin{equation}
\bigSigma_{kl}= \frac{1}{N} \sum_{i=1}^{N}{ ((\vvec_i)_k - {\langle \vvec 
\rangle}_k) ((\vvec_i)_l - {\langle \vvec \rangle}_l) }
\end{equation}
in the standard way.  $\Sigma$ is a 3 by 3 symmetric matrix, with $\Sigma_{12} = \mbox{Cov($x,y$)}$, $\Sigma_{13} = \mbox{Cov($x,m$)}$, and so on. 

\subsection{Estimate the joint density $p(\vvec)$}
\label{sec:jointDensity}

In Sec.~\ref{sec:generateMonteCarlo} we imagined filling a three-dimensional histogram in $\vvec$ with Monte Carlo events, and we recognized that the resulting histogram represents an estimation of a probability density.  A well-known technique in multivariate statistics involves estimating a probability density not by filling a histogram, but rather by summing kernels of probability placed around each point.  Due to its familiarity and smoothness properties, a favorite kernel choice is the multivariate gaussian:

\begin{equation}
\label{eqn:Kdef}
\bigK(\vvec) = \frac{1}{(\sqrt{2 \pi} h)^{3} \sqrt{\det(\bigSigma)}} 
\exp{ \left( \frac{ - \vvec^T \bigSigma^{-1} \vvec }{2 h^2} \right)}.
\end{equation}
The vector $\vvec$ is the same three-component vector defined above, and $\Sigma^{-1}$ is the inverse of the training array covariance matrix $\Sigma$.  The parameter $h$ is known in the language of density estimation as a {\em smoothing parameter}; it controls the width of the kernels placed around each point.  Theoretical arguments suggest an optimal choice of $h \approx N^{-1/(d+4)}$ as a function of the number of data points $N$ and the dimensionality $d$ of the variable space.\footnote{This expression for $h$ depends on assumptions about the probability density that we have not made explicit, and is not exact~\cite{Scott,Wand}.  In practice, $h$ may be optimized for any set of Monte Carlo events by constructing and minimizing some appropriate error estimate $\chi(h)$.  For $N=10^5$ and $d=3$, the optimal choice for $h$ is roughly 0.20.}    

An estimate of the joint probability density $p(\vvec)$ is then obtained simply by summing kernels centered about each of the $N$ data points $\vvec_i$, so that

\begin{equation}
\label{eqn:jointDensity}
p(\vvec)= \frac{1}{N} \sum_{i=1}^{N} { \bigK(\vvec - \vvec_i) }.
\end{equation}

\subsection{Compute the posterior density $p(m|\xvec)$}
\label{sec:posterior}

A physicist attempting to measure the top quark mass is interested in the {\em posterior density} $p(m|\xvec)$ for $m$.  In words, $p(m|\xvec)$ is the probability that the top quark mass is $m$ given that we have observed an event with variable values $\xvec$.  This posterior density is easily obtained.  The probability of obtaining both 
$\xvec$ and $m$ is equal to the probability of obtaining $\xvec$ multiplied by the 
probability of obtaining $m$ given that you have obtained $\xvec$:
\begin{equation}
p(\xvec,m)=p(\xvec)p(m|\xvec),
\end{equation}
and the probability of obtaining $\xvec$ is given by integrating the probability 
of obtaining both $\xvec$ and $m$ over all values of $m$:
\begin{equation}
p(\xvec)=\int{p(\xvec,m') \, dm'}.
\end{equation}
Thus the posterior density $p(m|\xvec)$ is related to the joint density 
$p(\xvec,m)$ simply by
\begin{equation}
\label{eqn:posteriorDensity}
p(m|\xvec)= \frac{p(\xvec,m)}{\int{p(\xvec,m') \, dm'}}.
\end{equation}

\subsection{Compute $\mhat$ (optional)}
\label{sec:bestestimate}

In the Bayesian view, the posterior density is the natural result of this recipe.  Nonetheless, it is often convenient to reduce the posterior density $p(m|\xvec)$ to a single number $\mhat$ representing the {\it best estimate} of the parameter in question.  Among the natural choices for the best estimate are the mean, median, and mode of the posterior distribution.  Adopting the last for the purposes of this discussion, we solve the equation 
\begin{equation}
\label{eqn:max}
p(\mhat|\xvec) = \max_m{p(m|\xvec)}
\end{equation}
numerically for $\mhat$.  Since the denominator of Eq.~\ref{eqn:posteriorDensity} is independent of $m$, maximizing the posterior density $p(m|\xvec)$ is equivalent to maximizing the joint density $p(\xvec,m)$, which we have constructed explicitly.  The extent to which the posterior density $p(m|\xvec)$ peaks around the value $\mhat$ depends, of course, on how strongly the variables $\xvec$ correlate with the true mass $m$.  

We note that this method can be modified to produce results that obey the frequentist notion of coverage.  Assume that a 68\% confidence region is desired.  Starting with $p(\xvec|m)$, draw for each fixed $m$ the contour ${\mathcal C}_m$ in $\xvec$-space enclosing 68\% of the density and minimal area.  Then upon observing $\xvec$ in the data, the 68\% confidence region for $m$ is the union of all values of $m$ for which $\xvec$ lies inside ${\mathcal C}_m$.

\vskip 1.0cm 

Extension to the case of an ensemble of data events is treated in Appendix~\ref{sec:bkgEvents}.

\section{Conclusions}
\label{sec:Conclusions}

The analysis method described here is quite general, and can be used in the context of any parameter estimation problem.  The non-parametric approach used to estimate probability densities is helpful when the distributions under consideration do not lend themselves to an obvious parameterization.  \alphaPDE\ allows the use of several measured variables, and enables the simultaneous estimation of several parameters.  The generalization to arbitrary dimension is provided in Appendix~\ref{sec:arbitraryDimension}.  Bayesian credible intervals and moments are easily obtained from simple manipulations of the joint probability density.

\appendix

\section{The general multivariate case}
\label{sec:arbitraryDimension}

For pedagogical reasons, \alphaPDE\ has been introduced through a specific 
example --- determining the mass $m$ of the top quark from two measured quantities $x$ and $y$ --- and the expressions in the text are therefore specific to that example.  In this appendix we provide the formulae for the general case.

In the general case, let each event be characterized by $d_1$ known variables $\xvec$ and $d_2$ unknown parameters $\avec$. Let $d=d_1+d_2$,  and let the $d$-dimensional event vector be $\vvec=(\xvec,\avec)$.

The $i^{\rm th}$ Monte Carlo event is now described by the event vector 
\begin{equation}
\vvec_i = (\xvec_i,\avec_i),
\end{equation}
 and the entire Monte Carlo sample is described by the training array 
\begin{equation}
T_{ij}=(\vvec_i)_j,
\end{equation}
 where $j$ now ranges from 1 to $d$.  The mean event vector is 
\begin{equation}
\langle \vvec \rangle = \frac{1}{N} \sum_{i=1}^{N}{\vvec_i}
\end{equation}
and the training covariance matrix is
\begin{equation}
\bigSigma_{kl}= \frac{1}{N} \sum_{i=1}^{N}{ ((\vvec_i)_k - {\langle \vvec 
\rangle}_k) ((\vvec_i)_l - {\langle \vvec \rangle}_l) },
\end{equation}
as before, and the general multivariate gaussian is given by 
\begin{equation}
\label{eqn:multiGaussian}
\bigK(\vvec) = \frac{1}{(\sqrt{2 \pi} h)^{d} \sqrt{\det(\bigSigma)}} 
\exp{ \left( \frac{ - \vvec^T \bigSigma^{-1} \vvec }{2 h^2} \right)}.
\end{equation}
Finally, in Eqs.~\ref{eqn:posteriorDensity} and~\ref{eqn:max}, $m$ should be replaced by the vector $\avec$.

In practice, limited computing resources place an upper bound on $N$, and hence an upper bound on $d$.  The optimal accuracy of the kernel estimate is of order $N^{-s/(2s+d)}$, where $s$ is a positive integer that reflects the assumed smoothness of the unknown density, and a typical assumption of continuous and square integrable derivatives up to second order corresponds to $s=2$~\cite{Scott,Wand}.

\section{Alternative to generating a random sample of Monte Carlo events}

In this appendix we describe a modification to the procedure described in the text if practical constraints prevent the generation of events pulled from a continuous prior $p(m)$, but allow the generation of events at $q$ discrete values $m_j$, where $j = 1, . . , q$.

Two changes are required in the first five steps of the recipe (Secs.~\ref{sec:prior}--\ref{sec:jointDensity}).  First, it is assumed that practical constraints require Monte Carlo events to be generated at the discrete masses $m_j$, rather than as described in Sec.~\ref{sec:generateMonteCarlo}.  Second, the function calculated in Eq.~\ref{eqn:jointDensity}, which may no longer be interpreted as a joint density, should be re-labeled. For lack of a better alternative, call it $\xi(\vvec)$.

We now add a step 5$\frac{1}{2}$ between Secs.~\ref{sec:jointDensity} and~\ref{sec:posterior}.  The function $\xi(\vvec)$ is clearly not an appropriate density.  If events have been generated assuming five different masses $m_j$, a graph of $\xi(\vvec)$ might appear as shown in Fig.~\ref{fig:fig1}.  We see that the density has ridges along the values of $m$ for which events have been generated, with corresponding valleys in the regions between these values.

{\dofig {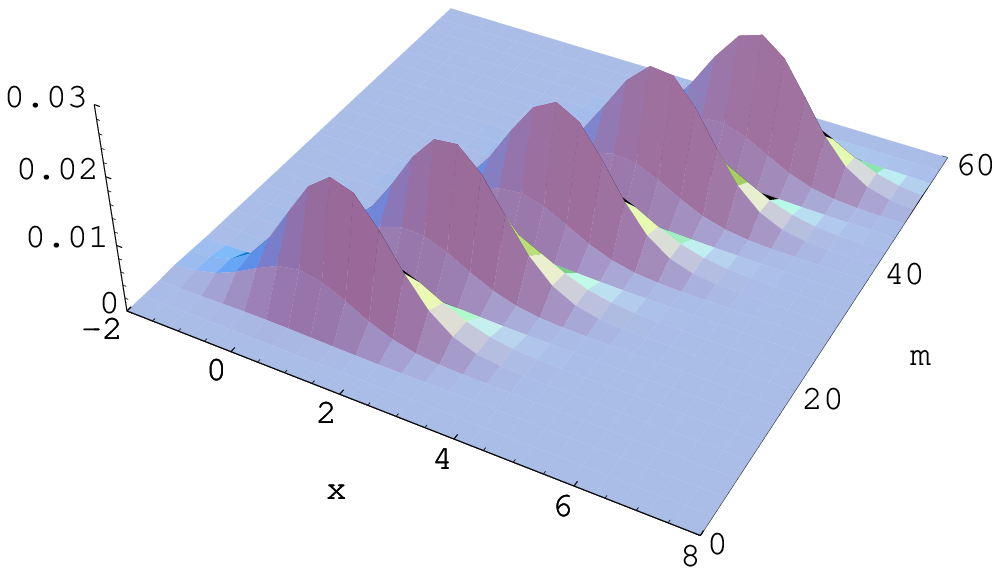} {3.0in} {A sample function $\xi(x,m)$ that might be constructed from Monte Carlo events at masses $m$ = 10, 20, 30, 40, and 50.  Notice the ridges in this function, due to the fact that it is constructed from events at specific masses.} {fig:fig1}}

An appropriately rescaled probability density $p(\xvec,m)$ can be generated by multiplying $\xi(\vvec)$ by a normalizing $m$-dependent factor $s(m)$:
\begin{equation}
p(\xvec,m)=\xi(\xvec,m) s(m).
\end{equation}
  This normalizing factor will correct for the fact that valleys have been introduced into the density by only generating events at specific masses $m_j$.  The requirement that
\begin{equation}
\int{p(\xvec,m) \, d\xvec} = p(m)
\end{equation}
determines this normalizing factor uniquely. 
The desired joint probability density $p(\vvec)$ is then given by
\begin{equation}
\label{eqn:gdef}
p(\vvec)= \frac{\xi(\vvec) p(m)}{ \int{d\xvec' \, \xi(\xvec',m) } },
\end{equation}
 and the final step (Sec.~\ref{sec:posterior}) is exactly as before.  The rescaled density of Fig.~\ref{fig:fig1} is shown in Fig.~\ref{fig:fig2}.

{\dofig {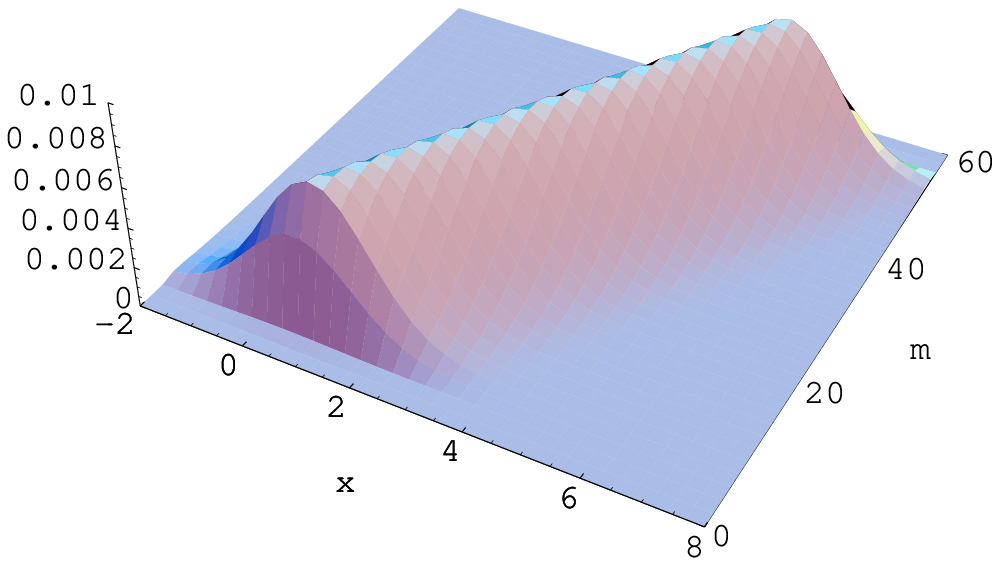} {3.0in} {The density $p(x,m)$ formed by rescaling the function $\xi(x,m)$ shown in Fig.~\ref{fig:fig1}.  Notice how this rescaling corrects for the fact that only events at specific masses were used in the construction of $\xi(x,m)$.} {fig:fig2}}

We mention briefly a useful shortcut when calculating integrals such as that appearing in the denominator of Eq.~\ref{eqn:gdef}.  Multidimensional integrals are difficult to calculate in general, but this integral can be handled analytically provided one uses gaussian kernels.  Assume as in Appendix~\ref{sec:arbitraryDimension} that the vector of known variables $\xvec$ is of $d_1$ dimensions, that the vector of unknown variables $\vec{\alpha}$ is of $d_2$ dimensions, and that the Monte Carlo has a covariance matrix $\Sigma$.  Then the relevant formula is 
\begin{equation}
\int{ K(\xvec,\vec{\alpha}) \, d\xvec} = 
	\frac{1}{(\sqrt{2 \pi} h)^{d_2} \sqrt{\det(\bigSigma')}} 
\exp{ \left( \frac{ - \vec{\alpha}^T \bigSigma'^{-1} \vec{\alpha} }{2 h^2} \right)},
\end{equation}
where $\bigSigma'$ is the $d_2$ by $d_2$ sub-matrix of $\bigSigma$ formed by retaining elements with row and column numbers larger than $d_1$.

\section{Background events}
\label{sec:bkgEvents}

In the text we considered the problem of determining the top quark mass $m$ for one candidate event.  In a real analysis there will be $n$ such events, and of those some fraction $b$ are expected to be background events --- events that do not contain a top quark at all.  This appendix shows how to apply \alphaPDE\ to a complete analysis.

Signal and background Monte Carlo events are generated and used to construct the signal density $p_s(\xvec,\avec)$, as described in Secs.~\ref{sec:prior}--\ref{sec:jointDensity}, and the background density $p_b(\xvec)$, which is independent of $\avec$.  From a careful analysis of background efficiencies we determine the probability $p(b)$ that a fraction $b$ of our events are background events.  In previous sections of this article $\xvec_i$ referred to Monte Carlo events; in this section we change notation and label the $n$ observed data events by $\xvec_1, .., \xvec_n$. 
 
The goal is to compute the posterior density $p(\avec|\xvec_1,.. ,\xvec_n)$.  Since the observations $\xvec_1,..,\xvec_n$ are assumed to be independent, $p(\xvec_1,..,\xvec_n|\avec,b)$ factors into a product:
\begin{equation}
\label{eqn:eqn15}
p(\xvec_1,..,\xvec_n|\avec,b)= \prod_{i=1}^n{p(\xvec_i|\avec,b)}.
\end{equation}
The probability $p(\xvec_i|\avec,b)$ for the $i^{\rm th}$ data event can be written in terms of the signal and background probability densities as
\begin{equation}
\label{eqn:eqn16}
p(\xvec_i|\avec,b)= \left( 1-b \right) p_s(\xvec_i|\avec) + b \, p_b(\xvec_i),
\end{equation}
where $p(\xvec|\avec)=p(\xvec,\avec)/p(\avec)$.  Integrating out the {\em nuisance parameter} $b$ in Eq.~\ref{eqn:eqn15} leaves
\begin{equation}
p(\avec|\xvec_1,..,\xvec_n) =  {\mathcal N} p(\vec{\alpha}) \int_0^1{ \left(\prod_{i=1}^n{\left[ \left( 1-b \right) p_s(\xvec_i|\avec) + b \, p_b(\xvec_i)\right]}\right) p(b)\,db},
\end{equation}
where ${\mathcal N}$ is a normalization factor ensuring that $\int{p(\avec|\xvec_1,..,\xvec_n) \, d\avec}=1$, and $p(\vec{\alpha},b)=p(\vec{\alpha})p(b)$ is assumed.

The most likely values of the parameters $\avec$ are then those for which $p(\avec|\xvec_1,..,\xvec_n)$ achieves its maximum, and the uncertainty on these values can be estimated from the width of the peak.  Other frequently-used best estimates and their errors are easily computed, if desired, from straightforward manipulation of the posterior density $p(\avec|\xvec_1,..,\xvec_n)$.

\bibliographystyle{plain}

\end{document}